\documentclass[10pt,conference]{IEEEtran}

\usepackage{amsmath,amssymb}
\usepackage{graphicx}
\usepackage{booktabs}
\usepackage{cite}
\usepackage{xcolor}

\title{Reducing quantum and classical resources for quantum-centric supercomputing workloads on near-term hardware}

\author{
\IEEEauthorblockN{Maxence Grandadam, Vladyslav Bohun, Dikshant Dulal, Maciej Koch-Janusz}
\IEEEauthorblockA{Haiqu, Inc., 95 Third Street, San Francisco, CA 94103, USA \\
\ maxence@haiqu.ai, vladbohun@haiqu.ai, dikshant@haiqu.ai, maciej@haiqu.ai}
}

\begin{document}
\maketitle

\begin{abstract}
Sample-based Krylov quantum diagonalization (SKQD) is a paradigmatic example of a quantum-centric supercomputing workflow that combines the convergence structure of Krylov quantum diagonalization with classical sampling-based post-processing. It provides convergence guarantees assuming that the important computational-basis configurations can be sampled from a set of Krylov states with sufficient probability. We analyze this assumption under depolarizing noise, deriving shot-count resource estimates that expose an exponential depth penalty, and perform experiments on current noisy hardware with the one-dimensional single-impurity Anderson model on a 20-site (40-qubit) instance. Device noise breaks the practical convergence predicted by the noiseless SKQD analysis, but approximate compilation techniques can compress the Krylov time-evolution circuits before execution. The compressed circuits accumulate less hardware noise, recover the expected energy convergence, and reduce both the quantum shot budget and the classical subspace dimension, remaining beneficial even when classical configuration recovery is applied.
\end{abstract}

\section{Introduction}
Estimating ground-state energies of correlated fermionic systems is a central target for quantum simulation~\cite{georgescu2014quantum}. Fully fault-tolerant approaches such as phase estimation remain too deep for present devices, while shallow variational algorithms face difficult optimization and measurement-scaling barriers~\cite{peruzzo2014vqe}. Subspace methods provide an intermediate route: the quantum processor prepares physically motivated states, and a classical computer diagonalizes the Hamiltonian projected onto the sampled or measured subspace~\cite{mcclean2017qse,epperly2022qsd}.

Sample-based Krylov quantum diagonalization (SKQD)~\cite{yu2025skqd} is particularly attractive for near-term hardware because it merges two complementary ideas. From Krylov quantum diagonalization, it inherits a convergence mechanism based on real-time evolution of an initial state~\cite{klymko2022realtime,shen2023realtime}. From sample-based quantum diagonalization (SQD), it inherits a noise tolerance crucial for near-term implementation by decoupling the sampling task, done by the quantum computer, from the evaluation of Hamiltonian matrix elements, done classically~\cite{robledo2024sqd}. This division of labor between a quantum sampler and classical post-processing is the essence of the quantum-centric supercomputing paradigm~\cite{bravyi2022future,alexeev2024qcsc}. In the ideal (noiseless) analysis, if the important computational-basis strings supporting the ground state are sampled, the final energy estimate converges according to the sparsity of the ground state in the chosen orbital basis.

This paper highlights a practical obstruction that arises in the presence of noise, but also shows that it can be efficiently mitigated. Using the one-dimensional single-impurity Anderson model (SIAM) as a prototypical example, we observe that hardware noise invalidates the useful part of the SKQD guarantee: the deeper Krylov circuits that should improve the subspace also accumulate more noise, leading to no new useful bitstrings being sampled. However, approximate compilation addresses this by replacing each time-evolution circuit with a shorter circuit that is close enough to the intended state but has substantially fewer noisy operations. The result is a better quantum/classical resource trade-off: lower energies are obtained with fewer shots and smaller classical subspaces. This advantage persists when the classical configuration-recovery mitigation is included, making approximate compilation an enabling step for scaling hybrid SKQD workloads.

%In the following, we start by presenting the theoretical framework on which SKQD is based (Sec.~\ref{sec:skqd}) and we extend it to the case with a depolarizing noise channel (Sec.~\ref{sec:noise}). We then present the idea of approximate compilation (Sec.~\ref{sec:compilation}) and how it fits into this picture. Lastly, this idea is used in practice with an experiment on current IBM hardware in Sec.~\ref{sec:experiment}.

\section{Sample-Based Krylov Quantum Diagonalization}
\label{sec:skqd}
SKQD builds a Krylov space by real-time evolution of a reference state and diagonalizes the Hamiltonian $H$ in the subspace spanned by the sampled computational-basis configurations. The Krylov states are generated from a reference state $|\psi_0\rangle$ by repeated time evolution,
\begin{equation}
\lvert \psi_k\rangle = e^{-i k \Delta t H} \lvert \psi_0 \rangle,
\quad k=0,\ldots,d-1,
\end{equation}
where $d$ is the Krylov dimension.
The quantum processor samples bitstrings $\lvert b \rangle$ in the computational basis from each $\lvert \psi_k \rangle$, and their union S defines a subspace of computational-basis states in which H is projected and diagonalized classically.

The SKQD guarantee rests on ground-state sparsity in the chosen basis and separates into two statements, only one of them being affected by noise. Write the ground state as $|\phi_0\rangle=\sum_b g_b|b\rangle$ and let $\mathcal{I}$ be a set of $L$ important bitstrings. We summarize the ground-state sparsity by the pair of parameters $(\alpha_0,\beta_0)$, where $\alpha_0=\sum_{b\in\mathcal{I}}|g_b|^2$ is the sum of the amplitudes of the important bitstrings, and $\beta_0=\min_{b\in\mathcal{I}}|g_b|^2$ is the smallest of their probabilities. The first number sets the achievable error, while the second is tied to the sampling cost.

On the one hand, if every bitstring in $\mathcal{I}$ is contained in the sampled set $\mathcal{S}$, the energy estimate obtained by diagonalizing $H$ in $\operatorname{span}(\mathcal{S})$ obeys~\cite{yu2025skqd}
\begin{equation}
\Delta E=\langle\Psi|H|\Psi\rangle-E_0\le 2\sqrt{2}\,\|H\|\left(1-\sqrt{\alpha_0}\right)^{1/2},
\label{eq:floor}
\end{equation}
where $E_0$ is the ground-state energy, $\|H\|$ the spectral norm of $H$, and $|\Psi\rangle$ the approximated ground state returned by the diagonalization. The achievable error is therefore fixed by $\alpha_0$ alone, a property of $H$ and independent of how the samples were produced. This is in sharp contrast to Krylov quantum diagonalization, where hardware noise corrupts the estimated pair $(\widehat H,\widehat S)$ and regularization induces a lower-bound on the achievable energy error~\cite{kirby2024krylov}. Here the projected problem is solved exactly on classically evaluated matrix elements, so there is no noise-induced lower-bound.

On the other hand, recovering all $L$ important bitstrings is where the circuits enter. For independent samples the recovery probability of a bitstring $b$ is
\begin{equation}
q_b = 1-\prod_{k=0}^{d-1}\left(1-p_k(b)\right)^M,
\quad
p_k(b)=|\langle b|\psi_k\rangle|^2,
\end{equation}
with $M$ shots per Krylov state. The construction guarantees~\cite{yu2025skqd} that every important bitstring has non-negligible weight in at least one state, $\max_k p_k(b)\ge p:=|\gamma_0|^2\beta(d)/d^2$, where $|\gamma_0|^2$ is the overlap between the reference and the ground state and $\beta(d)$ is the finite-$d$ sampling sparsity. At finite $d$ the Krylov states resolve the ground state only up to a geometrically small error, reducing the sparsity to $\beta(d)=\beta_0-B\,r^{-(d-1)/2}$, where $r=1+\pi\Delta E_1/\Delta E_{N-1}$ is the Krylov ratio set by the gap $\Delta E_1$ and spectral width $\Delta E_{N-1}$, and $B=2\sqrt{A/\Delta E_1}$ is the coefficient of the finite-$d$ correction, set by the amplitude $A\propto(1-|\gamma_0|^2)/|\gamma_0|^2$ of the geometric Krylov error, so that a larger reference overlap $|\gamma_0|^2$ (or gap $\Delta E_1$) reduces the correction. This sparsity approaches its intrinsic value, $\beta(d)\to\beta_0$, as $d\to\infty$. The union bound then guarantees $\Pr[\mathcal{I}\subseteq\mathcal{S}]\ge 1-\eta$, with target failure probability $\eta$, once the shots per Krylov state satisfy
\begin{equation}
M(d)\ge\frac{d^2\,\log(L/\eta)}{|\gamma_0|^2\bigl(\beta_0-B\,r^{-(d-1)/2}\bigr)}.
\label{eq:noiseless-bound}
\end{equation}
The $d^2$ prefactor is a polynomial cost, and the only other $d$-dependence is through $\beta(d)$. Without noise one increases $d$ until $\beta(d)$ saturates near $\beta_0$; because the gap closes geometrically this takes only logarithmic depth, leaving the overhead set by the ground-state sparsity.

\section{Noisy Sampling and the Depth Trade-Off}
\label{sec:noise}
\begin{figure}[t]
\centering
\includegraphics[width=.8\columnwidth]{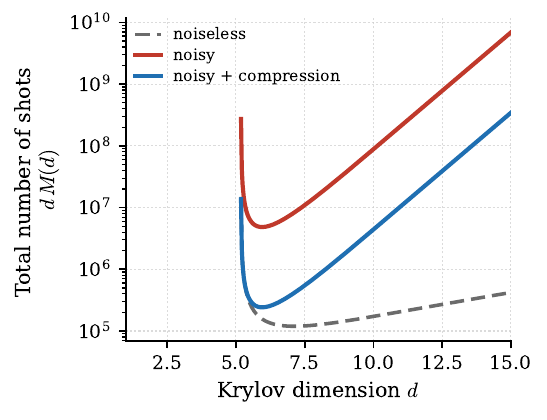}
\caption{Sufficient shots per Krylov state $M(d)$ from the support-recovery bound, in the three regimes (representative parameters). Without noise the cost grows only polynomially in $d$. Device noise multiplies it by $\lambda^{-(d-1)}$, an exponential penalty. Approximate compilation removes a fixed, $d$-independent number of noisy layers, lowering the requirement by the constant factor $G=(1-e)^{-(c_0+\ell c-c'_\ell)}$, so at any target Krylov dimension the same guarantee is met with $G$ times fewer shots per state. For this choice of parameters, $\beta(d) \le 0$ for $d \le 5$ and thus the bound in Eq.\eqref{eq:noiseless-bound} is not valid.}
\label{fig:theory}
\end{figure}
Noise changes the support-recovery probability before the classical diagonalization begins. Let $\lambda=(1-e)^{c}$ be the survival probability of one evolution step, for $c$ layers per step and depolarizing error $e$ per layer. A simple composition of global depolarizing channels for the circuit preparing the $k$-th Krylov state gives~\cite{nielsen2010quantum}
\begin{equation}
\widetilde\rho_k = \lambda^{k}\, |\psi_k\rangle\!\langle\psi_k| + \left(1-\lambda^{k}\right)\frac{I}{N},
\end{equation}
with $N$ the computational-basis dimension, and the observed probability of bitstring $b$ becomes
\begin{equation}
\widetilde p_k(b)=\lambda^{k}\, p_k(b)+\frac{1-\lambda^{k}}{N}.
\label{eq:noisy-prob}
\end{equation}
For an important bitstring, the uniform term should be negligible when $N$ is large, so its probability is reduced by the survival factor $\lambda^{k}$. Because only the existence of a suitable $k$ is guaranteed, the budget must cover the deepest circuit $k=d-1$, with survival $\lambda^{d-1}$. The sufficient shot count per Krylov state is then the noiseless bound~\eqref{eq:noiseless-bound} weighted by the survival of the deepest circuit,
\begin{equation}
M(d)\ge\frac{d^2\,\log(L/\eta)}{|\gamma_0|^2\,\lambda^{d-1}\bigl(\beta_0-B\,r^{-(d-1)/2}\bigr)}.
\label{eq:noisy-bound}
\end{equation}
The only change from the noiseless case is the factor $\lambda^{d-1}$ in the denominator, a penalty $\lambda^{-(d-1)}$ that grows exponentially with depth. There is now a trade-off when d increases between raising $\beta(d)$ toward $\beta_0$ and the additional cost given by $\lambda^{-(d-1)}$. This contrast is made explicit in Fig.~\ref{fig:theory}: without noise $M(d)$ grows only polynomially, whereas the factor $\lambda^{-(d-1)}$ makes it diverge exponentially, so a fixed shot budget already caps the affordable Krylov depth.

This balance is set by a single ratio,
\begin{equation}
\kappa=\frac{\ln(1/\lambda)}{\ln r}\approx\frac{c\,e\,\Delta E_{N-1}}{\pi\,\Delta E_1},
\label{eq:kappa}
\end{equation}
the infidelity accrued per evolution step ($c$ layers) per unit of spectral progress. Minimizing Eq.~\eqref{eq:noisy-bound}, to leading order in $1/d$ (dropping the $2/d$ contribution of the $d^2$ prefactor against the $O(1)$ noise term), gives an interior optimum in closed form,
\begin{equation}
d^\ast=1+\frac{2}{\ln r}\,\ln\!\frac{B\,(1+2\kappa)}{2\kappa\,\beta_0},\qquad \beta(d^\ast)=\frac{\beta_0}{1+2\kappa},
\label{eq:dstar}
\end{equation}
so the optimal sampling sparsity is reduced to a fraction $1/(1+2\kappa)$ of its converged value $\beta_0$ and the algorithm operates below full Krylov convergence. Because $c$ grows with circuit width and $\Delta E_{N-1}$ is extensive, $\kappa$ grows with system size, and reducing the number of noisy layers per state is the natural target for approximate compilation.

The same depth that suppresses the sampling probability of the important bitstrings also produces invalid samples. With $M$ shots per Krylov state, the expected number of nonideal samples is
\begin{equation}
S_{\rm fault}=M\sum_{k=0}^{d-1}\left(1-\lambda^{k}\right).
\label{eq:fault-samples}
\end{equation}
These are not merely wasted shots. Many become spurious bitstrings that enlarge the classical subspace, whose dimension $D$ controls the classical cost. Since Hamiltonian construction, configuration recovery, memory, and iterative diagonalization all grow with $D$, depth-induced noise is paid twice, first in the sampling bound and then in the classical post-processing.

\section{Approximate Compilation Advantage}
\label{sec:compilation}
Although there are many possible ways of approximately recompiling a quantum circuit in a shallower form, we focus here on the circuit compression technique developed by Haiqu, which is akin to the tensor-network approach to approximate quantum compilation~\cite{robertson2025aqc}. Accounting for the initial state preparation, the depth-$k$ circuit has $c_0+kc$ layers and survives with probability $(1-e)^{c_0}\lambda^{k}$. For each Krylov index we compress the initial state together with the first $\ell<k$ Trotter steps into a shallower circuit of measured fidelity $|\langle\varphi_\ell|\psi_\ell\rangle|^2=1-\delta_\ell$, leaving the remaining $k-\ell$ steps untouched. The horizon $\ell$ is set by entanglement growth, beyond which no low-depth block stays faithful. Because the tail of exact Trotter evolution is untouched, the infidelity is frozen at $\delta_\ell$ for every $k>\ell$, with no accumulation, and the deepest circuit now survives with probability $(1-e)^{c'_\ell}\lambda^{k-\ell}$, where $c'_\ell$ is the number of layers in the compressed block. The survival therefore improves by the $d$-independent factor $(1-e)^{-(c_0+\ell c-c'_\ell)}$.

A coherent mismatch cannot enter the sampling probabilities as a survival factor, so its worst-case effect is additive rather than multiplicative. Writing $|\varphi_\ell\rangle=\sqrt{1-\delta_\ell}\,|\psi_\ell\rangle+\sqrt{\delta_\ell}\,|\psi^\perp_\ell\rangle$, the probability of any bitstring obeys
\begin{equation}
\widehat p_k(b)\ge (1-e)^{c'_\ell}\lambda^{k-\ell}\left[\sqrt{(1-\delta_\ell)\,p}-\sqrt{\delta_\ell\,\mu_b}\right]_+^2,
\label{eq:compiled-prob}
\end{equation}
with $[x]_+=\max(x,0)$ and $\mu_b=|\langle b|\psi^\perp_\ell\rangle|^2$ the overlap of the residual with the computational basis state $\lvert b\rangle$. The only quantity that can reduce the signal is the alignment $\mu_b$ of the fixed residual with the important bitstrings. In practice, this residual is unbiased over a large support, so the penalty averages to a multiplicative survival $\widehat p_k(b)\ge(1-\delta_\ell)\,p$, exactly like additional stochastic noise.

This gain shifts the entire shot-cost curve down by the same $d$-independent factor, leaving the optimal depth $d^\ast$ unchanged (Fig.~\ref{fig:theory}). At the Krylov dimension required for a target accuracy, compression therefore reaches the same support-recovery guarantee with $(1-e)^{-(c_0+\ell c-c'_\ell)}$ times fewer shots per state, and a proportionally smaller total sampling cost, rather than by running a deeper circuit. It is beneficial whenever the noise it removes outweighs the infidelity it injects,
\begin{equation}
e\,(c_0+\ell c-c'_\ell)\ \gtrsim\ \delta_\ell,
\label{eq:advantage-condition}
\end{equation}
with an additional alignment penalty $2\sqrt{\delta_\ell\,\mu_b/p}$ in the adversarial worst case.

\section{Experiment: Single-Impurity Anderson Model}
\label{sec:experiment}
We benchmark our approximate compilation technique on the one-dimensional single-impurity Anderson model~\cite{anderson1961localized,hewson1993kondo},
\begin{equation}
H=H_{\rm imp}+H_{\rm bath}+H_{\rm hyb},
\end{equation}
with 
\begin{align}
H_{\rm imp} &= \epsilon_d(n_{d\uparrow}+n_{d\downarrow}) + U n_{d\uparrow}n_{d\downarrow},\\
H_{\rm bath} &=-t\sum_{j,\sigma}(c^\dagger_{j,\sigma}c_{j+1,\sigma}+c^\dagger_{j+1,\sigma}c_{j,\sigma}),\\
H_{\rm hyb} &= V\sum_{\sigma}(d^\dagger_{\sigma}c_{0,\sigma}+c^\dagger_{0,\sigma}d_{\sigma}),
\end{align}
on a 20-site ($n=40$ qubit) instance. We use $t=1$ as our energy reference and set $U=10$, $\epsilon_d=-U/2$ and $V=1$. Each Krylov state $\lvert \psi_k \rangle$ is obtained by using $k$ second-order Trotter steps~\cite{suzuki1990fractal} with a time-step $\Delta t=0.2$. Following the SKQD construction, we work in the orbital basis that best represents the ground state, obtained by diagonalizing the quadratic (bath) part of the Hamiltonian. Writing the evolution in this momentum basis increases both $\alpha_0$ and $\beta_0$, lowering the final energy error attainable and the sampling cost at once.

\begin{table}[h]
\centering
\caption{Circuit metrics for the largest SIAM Krylov circuit, $d=10$, with and without using Haiqu's approximate compilation. The circuits are transpiled for the layout of the \texttt{ibm\_fez} device.}
\label{tab:circuit}
\begin{tabular}{lrr}
\toprule
Metric & Compressed & Original \\
\midrule
Depth & 26 & 2413 \\
Two-qubit-gate depth & 4 & 619 \\
Two-qubit gates & 74 & 6840 \\
Fidelity & 0.941 & - \\
\bottomrule
\end{tabular}
\end{table}

\begin{figure}[h]
\centering
\includegraphics[width=\columnwidth]{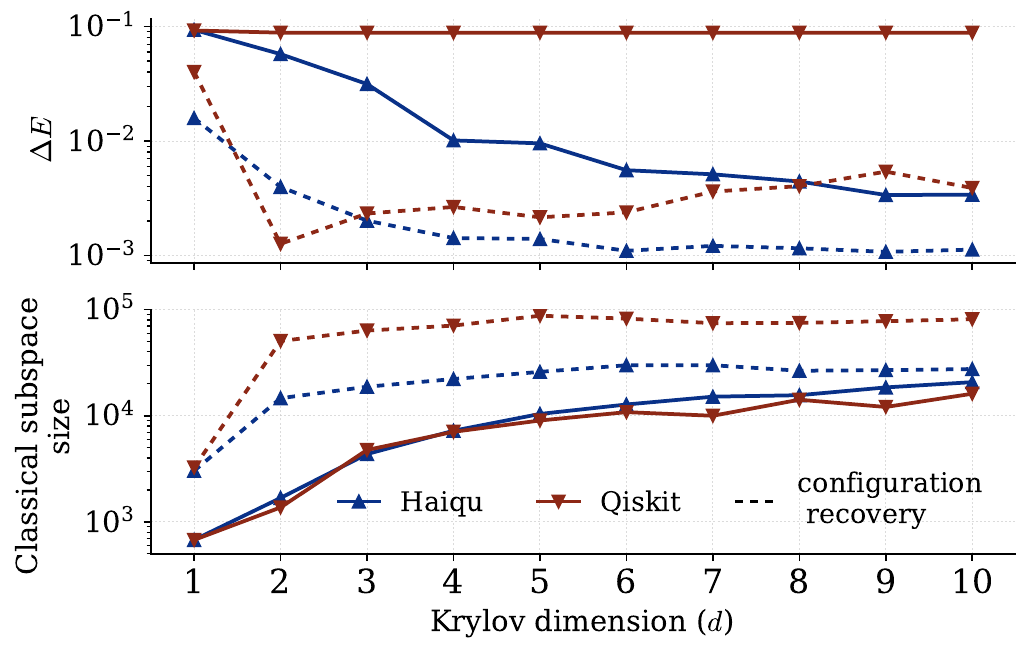}
\caption{Energy error $\Delta E$ (top) in units of the hopping parameter $t$ and dimension of the classical subspace (bottom) versus Krylov dimension $d$, for uncompressed (Qiskit) and compressed (Haiqu) circuits, each shown with and without configuration recovery. Samples are taken on \texttt{ibm\_fez} with $1000$ shots per Krylov state. The original Qiskit circuits stagnate near $\Delta E\approx9\times10^{-2}$ at every depth, dominated by noise, while the number of bitstrings included in the classical subspace keeps growing with $d$, a direct signature of the invalid samples of Eq.~\eqref{eq:fault-samples}. Approximate compilation restores monotonic convergence toward the reference ground-state energy. Reference ground-state energy is obtained by DMRG.}
\label{fig:convergence}
\end{figure}

Table~\ref{tab:circuit} shows that our largest compressed Krylov circuit reduces the total depth from $2413$ to $26$ layers, with the two-qubit gate count dropping from $6840$ to $74$, while achieveing a fidelity with respect to the original circuit of $\sim 0.94$. We note that this is a favorable case for the scheme: the one-dimensional geometry of the model keeps the entanglement bounded, so the entire Trotter evolution can be compressed into a single block. We sample the ten Krylov circuits ($k=0,\ldots,9$) on the \texttt{ibm\_fez} device with $1000$ shots per circuit (one circuit per Krylov state). From these samples we form the nested union sets $\mathcal{S}_i=\bigcup_{k<i}\mathcal{S}^{(k)}$, where $\mathcal{S}^{(k)}$ is the set of configurations sampled from Krylov state $k$, and for each $\mathcal{S}_i$ we run the classical iterative diagonalization with and without the configuration-recovery post-processing step, recording the final energy error and the dimension of the classical subspace. Fig.~\ref{fig:convergence} reports both quantities for the uncompressed and compressed circuits. The uncompressed Qiskit circuits stagnate far above the target energy, as expected when Eq.~\eqref{eq:noisy-bound} is dominated by the depth penalty. Haiqu's approximate compilation recovers monotonic improvement toward the reference ground-state energy while requiring a much smaller classical subspace.

\begin{figure}[t]
\centering
\includegraphics[width=\columnwidth]{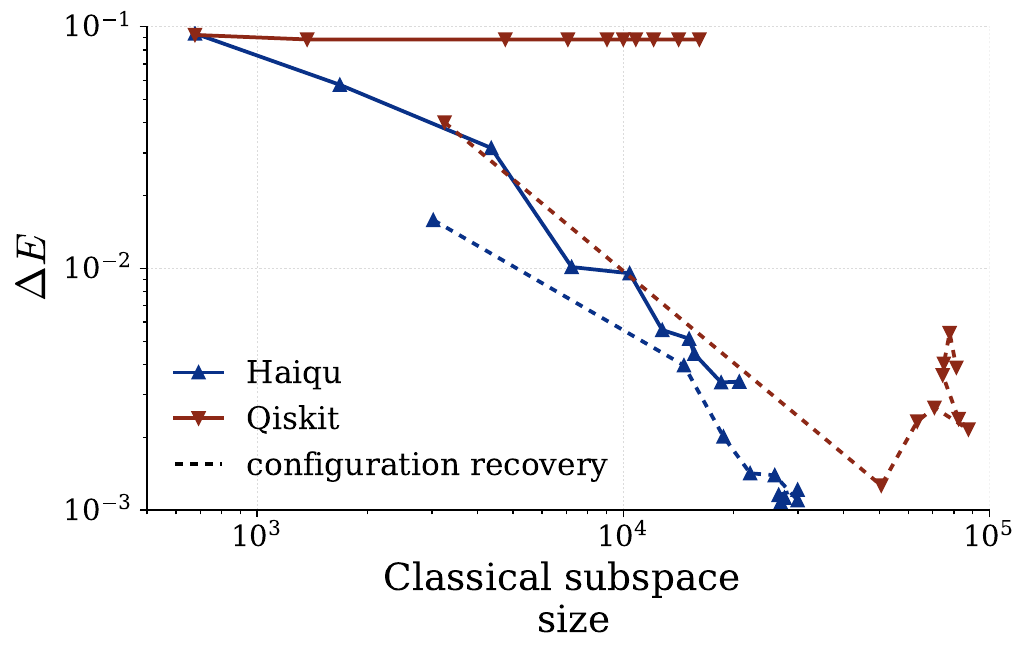}
\caption{Energy error $\Delta E$ versus classical subspace dimension, with and without configuration recovery. Each curve is the trajectory traced as the Krylov dimension increases from $d=1$ to $d=10$: uncompressed circuits (Qiskit) accumulate bitstrings without lowering the energy, whereas compressed circuits (Haiqu) show clear convergence with increased $d$. The best operating points combine approximate compilation with configuration recovery.}
\label{fig:pareto}
\end{figure}

Fig.~\ref{fig:pareto} shows the trade-off predicted by Eq.~\eqref{eq:fault-samples}. The best operating points are obtained by combining approximate compilation with configuration recovery. Configuration recovery remains valuable because it repairs samples using particle-number and occupancy information~\cite{robledo2024sqd,yu2025skqd}, but it cannot fully replace better quantum samples. Compressed circuits provide a cleaner starting distribution, which reduces the number of bitstrings that configuration recovery must generate before the classical diagonalization reaches a given energy, while providing a better initial averaged density for the iterative recovery~\cite{yu2025skqd}.

\section{Discussion and Conclusion}
We demonstrated how approximate compilation reduces the quantum and classical resource requirements of SKQD, a paradigmatic hybrid quantum-classical workload. While ground-state sparsity sets the minimum energy error obtainable, noise and compilation influence the price of reaching it. By raising the survival probability of each Krylov circuit our approach lowers the shot cost of support recovery and shrinks the determinant count $D$ which dominates the classical stage.  This picture is both supported by a theoretical analysis and borne out by experiments on noisy QPUs on a 20-site SIAM model, making approximate preparation of the Krylov states an important step toward large-scale SKQD workloads on hybrid quantum-HPC systems.

\section*{Acknowledgement}
M.G. acknowledges the use of Claude Opus 4.8 for text formatting, spell-checking and transcribing handwritten notes in LaTeX format.

\end{document}